\documentclass[twocolumn,preprintnumbers,secnumarabic,amsmath,amssymb,groupedaddress,nofootinbib]{revtex4}

\usepackage{graphicx}
\usepackage{dcolumn}
\usepackage{bm}
\usepackage{mathtools}

\newcommand{\boldmatrix}[1]{\bm{\mathit{#1}}}

\def\beq{\begin{equation}}
\def\eeq{\end{equation}}
\def\be{\begin{equation}}
\def\ee{\end{equation}}
\def\bea{\begin{eqnarray}}
\def\eea{\end{eqnarray}}

\begin{document}

\title{Vacua with Small Flux Superpotential}

\medskip\

\author{Mehmet Demirtas}
\email{md775@cornell.edu}
\affiliation{%
Department of Physics, Cornell University, Ithaca, NY 14853, USA}

\author{Manki Kim}%
\email{mk2427@cornell.edu}
\affiliation{%
Department of Physics, Cornell University, Ithaca, NY 14853, USA}

\author{Liam McAllister}%
\email{mcallister@cornell.edu}
\affiliation{%
Department of Physics, Cornell University, Ithaca, NY 14853, USA}

\author{Jakob Moritz}%
\email{moritz@cornell.edu}
\affiliation{%
Department of Physics, Cornell University, Ithaca, NY 14853, USA}

\date{\today}
\begin{abstract}

We describe a method for finding flux vacua of type IIB string theory in which the Gukov-Vafa-Witten superpotential is exponentially small.  We present an example with $W_0 \approx 2 \times 10^{-8}$ on an orientifold of a Calabi-Yau hypersurface with $(h^{1,1},h^{2,1})=(2,272)$, at large complex structure and weak string coupling.

\end{abstract}

\maketitle

\section{Introduction}
To understand the nature of dark energy in quantum gravity, one can study de Sitter solutions of string theory.
Kachru, Kallosh, Linde, and Trivedi (KKLT) have argued that there exist de Sitter vacua in compactifications on Calabi-Yau (CY) orientifolds of type IIB string theory \cite{Kachru:2003aw}.  An essential component of the KKLT scenario is a small vacuum value of the classical Gukov-Vafa-Witten \cite{Gukov:1999ya} flux superpotential,
\begin{equation}
W_0 := \sqrt{\tfrac{2}{\pi}}\,\Bigl\langle e^{\mathcal{K}/2}\int_X G \wedge \Omega \Bigr\rangle\,.
\end{equation}
Here $X$ is the CY orientifold, $G$ is the three-form flux, $\Omega$ is the $(3,0)$ form on $X$, $\mathcal{K}$
is the K\"ahler potential for the complex structure moduli and the axiodilaton, and the brackets denote evaluation on the vacuum expectation values of these moduli.  The stabilized values of the K\"ahler moduli are proportional to $\log( |W_0|^{-1})$, so control of the $\alpha'$ expansion is possible only if $|W_0|$ is very small.

String compactifications are characterized by discrete data, including the topology of the internal space, and quantized fluxes within it.  The number of distinct choices is vast, and although $|W_0| \ll 1$ is evidently not typical, strong evidence for the existence of vacua with $|W_0| \ll 1$ comes from the statistical treatment of
\cite{Ashok:2003gk,Denef:2004ze,Denef:2004cf,Douglas:2004zu,Douglas:2004kc}, as reviewed in \cite{Douglas:2006es}.
By approximating the integrally-quantized fluxes by continuous variables, one can compute the expected number of flux vacua with $|W_0| \le \delta$, for $\delta$ some chosen threshold.
This approach predicts that in an orientifold with a
sufficiently large value of the D3-brane charge tadpole $Q_{\mathrm{D3}}$ there should exist choices of flux giving vacua with exponentially small $|W_0|$.

We are not aware of any flaw in this statistical approach, but one can nevertheless ask: \emph{do there in fact exist} orientifolds and choices of flux giving vacua with $|W_0| \ll 1$, as the statistical theory predicts?  In this Letter we answer this question in the affirmative.

In \S\ref{sec:construction} we present a general method for constructing vacua with small $W_0$ at large complex structure (LCS) and weak string coupling, building on \cite{Giryavets:2003vd,Denef:2004dm}.
In \S\ref{sec:example} we give an explicit example\footnote{Pioneering work in this direction appears in \cite{Denef:2004dm,Denef:2005mm}.  Issues related to the size of $|W_0|$ are discussed in e.g.~\cite{Cicoli:2013swa,Palti:2019pca}.} where $W_0\approx 2 \times  10^{-8}$, in an orientifold of a Calabi-Yau hypersurface in $\mathbb{CP}_{[1,1,1,6,9]}$.  In \S\ref{sec:statistics} we show that our result accords well with the statistical predictions of \cite{Denef:2004ze}. We show in \S\ref{sec:fullstab} that at least one complex structure modulus in our example is as light as the K\"ahler moduli. We explain why this feature occurs in our class of solutions, and we
comment on K\"ahler moduli stabilization in our vacuum.

\section{A landscape of weakly coupled flux vacua with small $W_0$}\label{sec:construction}

Vacua with $|W_0| \ll 1$ are rare elements in a large landscape.
It is therefore impractical to exhibit vacua with $|W_0| \ll 1$ by enumerating general vacua on a massive scale and filtering out the desired ones. Instead one should pursue algorithms that preferentially find fluxes that lead to vacua with small $|W_0|$.

One algorithm of this sort\footnote{For an approach via genetic algorithms see \cite{Cole:2019enn}.} \cite{Giryavets:2003vd,Denef:2004dm} proceeds by finding quantized fluxes that solve an \emph{approximate} form of the F-term equations, with the corresponding approximate superpotential exactly vanishing, at some given point $U_{\star}$ in moduli space.  One then solves for nearby moduli values $U=U_{\star}+\delta U$ that solve the \emph{true} F-term equations with the same choice of fluxes.  When the approximation made in the first step is a good one, the true superpotential evaluated at $U=U_{\star}+\delta U$ will be small.

We will construct a class of flux vacua along these lines.  The approximate superpotential is obtained by neglecting nonperturbative corrections to the prepotential for the complex structure moduli around the LCS locus in moduli space.\footnote{Recent discussions of flux potentials near LCS appear in \cite{Honma:2017uzn,Grimm:2019ixq,Honma:2019gzp}.}
Stabilization near LCS, where these nonperturbative terms are exponentially small, then yields an exponentially small flux superpotential.

We consider an orientifold $X$ of a Calabi-Yau threefold with $-Q_{\mathrm{D3}}$ units of D3-brane charge on seven-branes and O3-planes.
Let $\{ A_a,B^b\}$ be a symplectic basis for $H_3(X,\mathbb{Z})$, with $A_a\cap A_b=0,$ $A_a\cap B^b=\delta^{~b}_a,$ and $B^a\cap B^b=0.$
We use projective coordinates $\{U^a\}$ on the complex structure moduli space of dimension $n\equiv h^{2,1}_-$, and we work in a gauge in which $U^0=1.$
Denoting the prepotential by $\mathcal{F}$ and writing $\mathcal{F}_a=\partial_{U^a}\mathcal{F},$ we define the period vector as
\begin{equation}
\Pi=\left( \begin{array}{c} \int_{B^a}\Omega\\ \int_{A_a}\Omega\end{array}\right)=\left(\begin{array}{c}\mathcal{F}_a\\ U^a \end{array}\right)\,.
\end{equation}
The integer flux vectors $F$ and $H$ are similarly obtained from the three-form field strengths $F_3$ and $H_3$ as
\begin{equation}
F=\left(\begin{array}{c}\int_{B^a}F_3\\ \int_{A_a}F_3 \end{array}\right), \qquad H=\left(\begin{array}{c}\int_{B^a}H_3\\ \int_{A_a}H_3 \end{array}\right)\,.
\end{equation}
Defining the symplectic matrix
$\Sigma=\begin{psmallmatrix}0 & I\\-I & 0\end{psmallmatrix}$,
the flux superpotential and the K\"ahler potential are\footnote{We have set the reduced Planck mass to unity, and
we omit here the K\"ahler potential for the K\"ahler moduli, which reads
$\mathcal{K}_K = -3\log\bigl(2\,\mathrm{Vol}^{2/3}\bigr)$, with $\mathrm{Vol}$ the volume of $X$ in ten-dimensional Einstein frame, in units of $(2\pi)^2\alpha'$.  Our conventions match those of \cite{Kachru:2019dvo} upon taking $\mathcal{V}=1/{4\pi}$ and $b=1$ in \S A.3 of \cite{Kachru:2019dvo}, cf.~\cite{Denef:2005mm}.}
\begin{align}\label{eq:wis}
W=&\,\sqrt{\tfrac{2}{\pi}}\Bigl(F-\tau H\Bigr)^T\cdot\Sigma \cdot\Pi \,, \\
\mathcal{K}=-&\log\Bigl(-i\Pi^{\dagger}\cdot\Sigma\cdot \Pi\Bigr)-\log\Bigl(-i(\tau-\bar{\tau})\Bigr)\,.
\end{align}
The LCS expansion of the prepotential is $\mathcal{F}(U)= \mathcal{F}_{\text{pert}}(U)+\mathcal{F}_{\text{inst}}(U)$
\cite{Hosono:1994av}  with the perturbative terms
\begin{equation}\label{eq:LCSprepotential}
\mathcal{F}_{\text{pert}}(U)=-\frac{1}{3!}\mathcal{K}_{abc}U^aU^bU^c+\frac{1}{2}a_{ab}U^aU^b+b_aU^a+\xi\, ,
\end{equation}
and the instanton corrections
\begin{equation} \mathcal{F}_{\text{inst}}(U)=\frac{1}{(2\pi i)^3}\sum_{\vec{q}}A_{\vec{q}}\,e^{2\pi i \vec{q}\cdot \vec{U}}\,.
\end{equation}
Here $\mathcal{K}_{abc}$ are the triple intersection numbers of the mirror CY, $a_{ab}$ and $b_a$ are rational,
the sum runs over effective curves in the mirror CY, and $\xi=-\frac{\zeta(3)\chi}{2(2\pi i)^3}$, with $\chi$ the Euler number of the CY.
We write
\begin{equation}
W=W_{\text{pert}}+W_{\text{inst}}\,,
\end{equation}
with $W_{\text{pert}}$ the portion obtained by using $\mathcal{F}_{\text{pert}}(U)$ in (\ref{eq:wis}), and $W_{\text{inst}}$
the correction from $\mathcal{F}_{\text{inst}}(U)$.
We call $W_{\text{pert}}$ the perturbative superpotential, and $W_{\text{inst}}$ the nonperturbative correction, even though the full flux superpotential $W$ is classical in the type IIB theory.

The real parts of $\vec{U}$ are axionic fields that do not appear in the perturbative K\"ahler potential, enjoying discrete gauged shift symmetries $\vec{U}\mapsto \vec{U}+\vec{\nu}$ with $\vec{\nu} \in \mathbb{Z}^n$. Under such a shift,
the period and flux vectors undergo a monodromy transformation $\{\Pi,F,H\}\mapsto M_{\infty}^{\vec{\nu}}\{\Pi,F,H\}$ with the monodromy matrix $M_{\infty}^{\vec{\nu}}\in Sp(2n+2,\mathbb{Z})$.
For generic choices of flux quanta, these discrete axionic shift symmetries are spontaneously broken, realizing axion monodromy \cite{Silverstein:2008sg,McAllister:2008hb,Kaloper:2008fb}.
A discrete shift symmetry remains unbroken
if and only if there exists a monodromy transformation $M_{\infty}^{\vec{\nu}}$ combined with an $SL(2,\mathbb
Z)$ transformation $T^r:(H,F)\mapsto (H,F+rH)$, $r\in \mathbb
Z$, that leaves the pair of flux vectors invariant.

Consider a choice of fluxes and moduli values that solves the F-flatness conditions, has an unbroken shift symmetry, and has
$W_{\text{pert}}=0$,
all at the level of the perturbative prepotential $\mathcal{F}_{\text{pert}}(U)$.
We call such a configuration a \textit{perturbatively flat vacuum}.
Here is a sufficient condition for the existence of such a vacuum.

\textbf{Lemma:} \textit{Suppose there exists a pair $(\vec{M},\vec{K})\in \mathbb{Z}^n\times \mathbb{Z}^n$ satisfying $-\frac{1}{2}\vec{M}\cdot \vec{K}\leq Q_{\mathrm{D3}}$ such that $N_{ab}\equiv \mathcal{K}_{abc} M^c$ is invertible, and $\vec{K}^T\boldmatrix{N}^{-1}\vec{K}=0$, and $\vec{p}\equiv \boldmatrix{N}^{-1}\vec{K}$ lies in the K\"ahler cone of the mirror CY, and such that $\boldmatrix{a}\cdot \vec{M}$ and $\vec{b}\cdot \vec{M}$ are integer-valued.  Then
there exists a choice of fluxes, compatible with the tadpole bound set by $Q_{\mathrm{D3}}$, for which a perturbatively flat vacuum exists.
The perturbative F-flatness conditions obtained from (\ref{eq:LCSprepotential}) are then satisfied along the one-dimensional locus $\vec{U}=\tau \vec{p}$ along which
$W_{\text{pert}}$ vanishes.}

To verify the Lemma, one considers the fluxes
\begin{equation}\label{eq:homogeneousfluxes}
F=(\vec{M}\cdot \vec{b},\vec{M}^T\cdot \boldmatrix{a},0,\vec{M}^T)\, ,\quad H=(0,\vec{K}^T,0,0)\, ,
\end{equation}
which can be shown to be the most general ones leading to a perturbative superpotential $W_{\text{pert}}$ that is homogeneous of degree two in the $n+1$ moduli.
The monodromy transformation $M_{\infty}^{\vec{\nu}}$ combined with an appropriate
$SL(2,\mathbb{Z})$ transformation leaves \eqref{eq:homogeneousfluxes} invariant, so a discrete shift symmetry
remains unbroken.

Because $W_{\text{pert}}$ is homogeneous, there is a perturbatively-massless modulus corresponding to an overall rescaling of the moduli.  This modulus can be stabilized by the nonperturbative terms in $\mathcal{F}$. Given $(\vec{M},\vec{K})$ for which stabilization of the rescaling mode occurs at weak coupling,
$W_{\text{inst}}$
will be exponentially small.
One finds the effective superpotential
\begin{equation}
\frac{W_{\mathrm{eff}}(\tau)}{\sqrt{2/\pi}}=M^a\partial_a \mathcal{F}_{\mathrm{inst}}=\sum_{\vec{q}} \frac{A_{\vec{q}}\,\vec{M}\cdot \vec{q}}{(2\pi i)^2}e^{2\pi i \tau  \vec{p}\cdot \vec{q}}\, ,
\end{equation}
where we have chosen the axiodilaton $\tau$ as a coordinate along the flat valley.
As the inner product $\vec{p}\cdot \vec{q}$ need not be integer, it is possible to find flux quanta such that $\tau$ can be stabilized at weak coupling, by realizing a racetrack.\footnote{Achieving racetrack stabilization within our class of models could aid in the search for large axion decay constants via alignment, as in \cite{Kim:2004rp}. See \cite{Hebecker:2015rya,Blumenhagen:2016bfp,Hebecker:2017lxm} for approaches using shift symmetries: in particular, \cite{Hebecker:2017lxm} also employs a superpotential of degree two.}  This works efficiently if the two most relevant instantons, which we label as $\vec{q}_1$ and $\vec{q}_2$, satisfy $\vec{p}\cdot \vec{q_1}\approx \vec{p}\cdot \vec{q_2}$.

\section{An example}\label{sec:example}

We now illustrate our method in an explicit example, with $n=2$.
We consider the degree 18 hypersurface in weighted projective space $\mathbb{CP}_{[1,1,1,6,9]}$ studied in \cite{Candelas:1994hw}. This is a CY with $272$ complex structure moduli, but as explained in \cite{Giryavets:2003vd}, it is useful to study a particular locus in moduli space where the CY becomes invariant under a $G=\mathbb{Z}_6\times \mathbb{Z}_{18}$ discrete symmetry.\footnote{If we were to orbifold by this symmetry group and resolve the singularities we would obtain the mirror CY \cite{Greene:1990ud}. We will \textit{not} proceed in this direction, but instead stay with the original CY.} By turning on only flux quanta invariant under $G$, we are guaranteed to find solutions of the full set of F-term equations, by solving only those corresponding to the directions tangent to the invariant subspace. Conveniently, the periods in these directions are identical to the periods of the mirror CY, and are obtained from an effective two-moduli prepotential as in \eqref{eq:LCSprepotential} with the data
\begin{align}\label{eq:geodata}
\mathcal{K}_{111}=&\,\,9\, ,\quad \mathcal{K}_{112}=3\, ,\quad \mathcal{K}_{122}=1\, ,\nonumber\\ \boldmatrix{a}=&\,\,\frac{1}{2}\begin{pmatrix}
9 & 3 \\
3 & 0
\end{pmatrix}\, ,\quad \vec{b}=\frac{1}{4}\begin{pmatrix}
17\\
6
\end{pmatrix}\, .
\end{align}
The instanton corrections take the form \cite{Candelas:1994hw}
\begin{align}\label{eq:theinst}
& (2\pi i)^3\mathcal{F}_{\text{inst}}=\mathcal{F}_1 + \mathcal{F}_2 + \cdots\, , \\
& \mathcal{F}_1 =  - 540 q_1 - 3q_2\, ,\label{eqf1} \\
& \mathcal{F}_2 = -\frac{1215}{2} q_1^2 +1080 q_1 q_2 + \frac{45}{8} q_2^2\, ,\label{eqf2}
\end{align}
where $q_i=\exp(2\pi i U^i)$ with $i \in \{1,2\}$. Note the $\mathcal{O}(10^{-2})$ hierarchy in the coefficients of the one-instanton terms.  We consider an orientifold involution described in \cite{Louis:2012nb},
with two O7-planes, each with four D7-branes, and in which we find
$Q_{\mathrm{D3}}=138.$

We will now use the Lemma to find a pair $(\vec{M},\vec{K})\in \mathbb{Z}^2\times \mathbb{Z}^2$
yielding a perturbatively flat vacuum.  Using \eqref{eq:geodata}, the condition $\vec{K}^T \boldmatrix{N}^{-1} \vec{K} = 0$ becomes
\begin{equation}\label{eq:lemmaeq}
	M_1 = \frac{M_2 K_2 (2K_1 - 3K_2)}{(K_1 - 3K_2)^2}\,,
\end{equation}
and the flat direction
is given by
\begin{equation}
	\vec{U} = \tau \begin{pmatrix}
	p_1\\
	p_2
	\end{pmatrix} = \frac{\tau (K_1 - 3K_2)}{M_2} \begin{pmatrix}
	- K_2/K_1\\
	1
	\end{pmatrix}\,.
\end{equation}
Once the nonperturbative corrections \eqref{eq:theinst} are included, the effective superpotential along the flat direction reads
\begin{equation}\label{weffis}
	W_{\text{eff}}(\tau)=c\,\Bigl(e^{2\pi i p_1 \tau}+Ae^{2\pi i p_2 \tau}\Bigr)\, + \cdots\, ,
\end{equation}
where $c$ and $A$ depend on the pair $(\vec{M}, \vec{K})$, but not on $\tau$. A racetrack potential is realized when the two terms in \eqref{weffis} are of comparable magnitude, which requires $|p_1-p_2| \ll p_2$. We are thus looking for $\vec{M}$ and $ \vec{K}$ obeying \eqref{eq:lemmaeq} for which $Q_{\mathrm{D3}}^{\mathrm{flux}} \equiv - \frac{1}{2} \vec{M} \cdot \vec{K} \leq 138$ and $|K_1+K_2| \ll |K_2|$.  A suitable choice is
\begin{equation}
\vec{M}=\begin{pmatrix*}
-16\\
\phantom{-}50
\end{pmatrix*}\, ,\quad \vec{K}=\begin{pmatrix*}
\phantom{-}3\\
-4
\end{pmatrix*}\, ,
\end{equation}
which gives $Q_{\mathrm{D3}}^{\mathrm{flux}}=124$, $A=-\frac{5}{288}$, and $c=-\sqrt{\tfrac{2}{\pi}}\frac{8640}{(2\pi i)^3}$. The resulting racetrack potential is depicted in Figure 1. The moduli are stabilized at
\begin{equation}
\langle \tau\rangle =6.856 i\, , \quad
\langle U_1\rangle =2.742 i\, , \quad
\langle U_2\rangle = 2.057 i\, ,
\end{equation}
and we find
\begin{equation}\label{theans}
|W_0|=2.037\times 10^{-8}\, .
\end{equation}
\begin{figure}
	\centering
	\includegraphics[keepaspectratio,width=7cm]{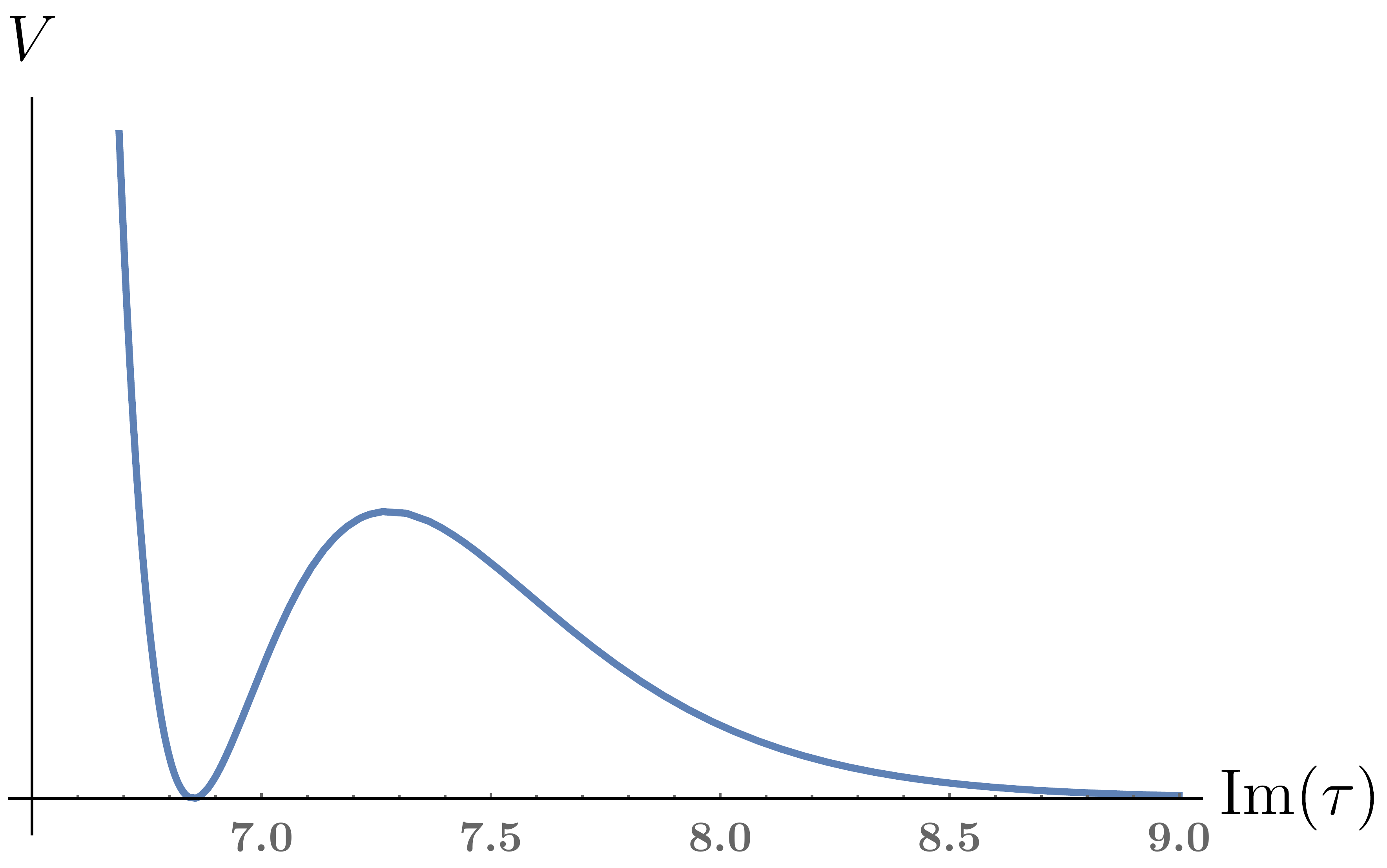}
	\caption{The scalar potential on the positive $\mathrm{Im}\,\tau$ axis.}
	\label{fig:racetrackpot}
\end{figure}
The instanton expansion is under excellent control: the two-instanton terms (\ref{eqf2}) are a factor $\mathcal{O}(10^{-5})$ smaller than the one-instanton terms (\ref{eqf1}), and the three-instanton terms are smaller by a further factor $\mathcal{O}(10^{-5})$.

\section{Statistics of small $W_0$}\label{sec:statistics}

A systematic understanding of statistical predictions of the flux landscape was developed in \cite{Ashok:2003gk,Denef:2004ze,Denef:2004cf,Douglas:2004zu,Douglas:2004kc}, in part by approximating the integer fluxes by continuous variables. Let us compare our result (\ref{theans}) with the statistical prediction for the smallest possible $W_0$ in our orientifold.

We write $\mathcal{N}(\lambda_*,Q_{\mathrm{D3}})$ for the expected number of vacua with D3-brane charge in fluxes less than $Q_{\mathrm{D3}}$ and with $|W_0|^2\leq \lambda_*\ll 1$.
According to \cite{Denef:2004ze}, $\mathcal{N}(\lambda_*,Q_{\mathrm{D3}})$ is given for $n=2$ by
\begin{equation}
\mathcal{N}(\lambda_*,Q_{\mathrm{D3}})=\frac{2\pi^4(2 Q_{\mathrm{D3}})^{5}}{5!}\lambda_*\int_{\mathcal{M}}
\star\, e^{2\mathcal{K}}\mathcal{F}_{abc}\overline{\mathcal{F}}^{abc}\,,\label{eqn:apx number of vacua small w0}
\end{equation}
where $\mathcal{M}$ is the axiodilaton and complex structure moduli space, $\star$ is the Hodge star on $\mathcal{M}$, and $\mathcal{F}_{abc}\equiv \partial^3_{abc}\mathcal{F}$. Taking $Q_{\mathrm{D3}}=138$ and numerically integrating over the LCS region $1<\text{Im}(U)$, we find that
$\mathcal{N}(\lambda_*,138)<1$ for $\sqrt{\lambda_*}\lesssim 6\times 10^{-7}$.  The prediction of \cite{Denef:2004ze} is thus that the smallest value of $|W_0|$ expected to exist is of order $6\times 10^{-7}$, which agrees reasonably well with (\ref{theans}).

\section{Toward stabilizing all moduli}\label{sec:fullstab}

Thus far we have found a class of no-scale vacua in which the complex structure moduli and axiodilaton F-terms vanish, and $W_0$ is exponentially small.
To achieve stabilization of the K\"ahler moduli from this promising starting point,
two issues must be addressed: the masses of the complex structure moduli and axiodilaton, and the nonperturbative superpotential for the K\"ahler moduli.

For the example of \S\ref{sec:example}, we have computed the mass matrix along the $G$-symmetric locus.  Two of the moduli are heavy, but the third, corresponding to the perturbatively-flat direction $\tau$, has a mass proportional to $|W_0|$.
We are not aware of a reason why any of the $G$-breaking combinations should be comparably light, but checking this directly will be important, and rather challenging.
Assuming that the $G$-breaking moduli are indeed heavy, the low energy theory describing K\"ahler moduli stabilization will include $\tau$ and the K\"ahler moduli $T_1$, $T_2$.
(K\"ahler moduli stabilization in a related setting has been discussed in \cite{Hebecker:2018fln}.)

Provided that the seven-brane stacks wrap divisors that are either rigid \cite{Witten:1996bn}, or else are rigidified by the introduction of fluxes \cite{Kallosh:2005yu,Kallosh:2005gs,Bianchi:2011qh}, we expect a nonperturbative superpotential of the form
\begin{align}\label{wtotis}
W_{\mathrm{eff}}(\tau,T_1,T_2)=&\,\,c\,\Bigl(e^{2\pi i \frac{2}{5} \tau}
+Ae^{2\pi i\frac{3}{10}\tau}\Bigr)\nonumber\\&+Be^{-\frac{2\pi }{c_1}T_1}+Ce^{-\frac{2\pi }{c_2}T_2}\,.
\end{align}
Here $A$ and $c$ are known coefficients, cf.~\eqref{weffis}, and $c_1$ and $c_2$ are the dual Coxeter numbers of the confining seven-brane gauge groups.
The unbroken discrete shift symmetry implies that the Pfaffian prefactors $B$ and $C$ can be expanded in appropriate powers of $e^{2\pi i \tau}$.  Provided that there exists an interpolation to a weakly curved type IIA description, these powers should be \emph{nonnegative}, because the objects that break the continuous shift symmetry to a discrete one are type IIB D(-1)-brane instantons, and worldsheet instantons of the type IIA mirror, which are negligible at small string coupling and large complex structure.  By neglecting the exponentially small corrections, one should then be able to treat $B$ and $C$ as constants.  Verifying this directly would be informative.

To exhibit vacua with all moduli stabilized in this setting, one should establish \eqref{wtotis}
and compute $B$ and $C$ for a seven-brane configuration in which $c_1$ and $c_2$ are sufficiently large to ensure control of the $\alpha'$ expansion.  This worthy goal is beyond the scope of the present work.
\vspace{0.5cm}

\section{Conclusions}  We have described a method for constructing flux vacua with exponentially small Gukov-Vafa-Witten superpotential in compactifications of type IIB string theory on Calabi-Yau orientifolds, at weak string coupling and large complex structure.  The first step is to neglect nonperturbative terms in the prepotential expanded around large complex structure, and find quantized fluxes that at this level yield vanishing F-terms and vanishing superpotential along a flat direction in the complex structure and axiodilaton moduli space.
We provided simple and constructive sufficient conditions for the existence of such solutions, and we determined the flat direction analytically, vastly simplifying the search for vacua.
Upon restoring the nonperturbative corrections, one can find full solutions in which the flat direction is lifted, although it remains anomalously light, and the flux superpotential is exponentially small.

We gave an explicit example with $|W_0|\approx 2\times 10^{-8}$ in an orientifold of the Calabi-Yau hypersurface in $\mathbb{CP}_{[1,1,1,6,9]}$.
This value of $|W_0|$ accords well with the statistical expectation derived from the work of Denef and Douglas \cite{Denef:2004ze}.
Stabilizing the K\"ahler moduli in this class of vacua, and then pursuing more explicit de Sitter solutions, are important tasks for the future.

{\bf Acknowledgements.}\
We thank Mike Douglas, Naomi Gendler, Arthur Hebecker, Shamit Kachru, Daniel Longenecker, John Stout, Irene Valenzuela, and Alexander Westphal for useful discussions, and we are grateful to Shamit Kachru, Daniel Longenecker, and Alexander Westphal for comments on a draft.  We thank Arthur Hebecker for correspondence about \S\ref{sec:fullstab}.
The work of M.D., M.K., and L.M.~was supported in part by NSF grant PHY-1719877, and the work of L.M.~and J.M.~was supported in part by the Simons Foundation Origins of the Universe Initiative.  M.K.~gratefully acknowledges support from a Boochever Fellowship.

\bibliography{refs}

\end{document}